\documentclass[journal=jacsat,manuscript=article]{achemso}

\usepackage{chemformula}
\usepackage[T1]{fontenc}
\usepackage{url}
\usepackage{braket}
\usepackage{booktabs}
\usepackage[version=4]{mhchem}
\usepackage{subcaption} 
\usepackage{enumitem}

\author{Daniel Graf}
\affiliation[University of Cambridge]
{Yusuf Hamied Department of Chemistry, University of Cambridge, Cambridge}
\alsoaffiliation[University of Munich]
{Department of Chemistry, Ludwig-Maximilians-Universit\"at M\"unchen, Munich}
\email{daniel.graf@cup.uni-muenchen.de}
\author{Alex J. W. Thom}
\affiliation[University of Cambridge]
{Yusuf Hamied Department of Chemistry, University of Cambridge, Cambridge}

\title[NOCI]
  {Natural Orbital Non-Orthogonal Configuration Interaction}


\begin{document}

\begin{abstract}
    Non-orthogonal configuration interaction (NOCI) is a generalization of the
    standard orthogonal configuration interaction (CI) method and offers a
    highly flexible framework for describing ground and excited electronic
    states. However, this flexibility also comes with challenges, as there is 
    still no clear or generally accepted approach for constructing a compact and
    accurate state basis for NOCI.
    In this work, we take a step toward addressing this challenge by introducing 
    a novel NOCI approach designed with three primary objectives: (1) ensuring
    the method is systematic, (2) achieving a compact NOCI expansion, and (3)
    treating all electronic states of interest on equal footing.
    The development of our approach is presented step by step, with each 
    building block evaluated and validated through applications to simple model 
    systems, demonstrating its effectiveness and potential.
\end{abstract}

\section{Introduction}
Excited states lie at the core of many fundamental processes in
nature, such as photosynthesis\cite{cerullo2002a, cheng2009a}, human
vision\cite{herbst2002a, tapavicza2007a}, photoinduced damage of essential
biomolecules\cite{schultz2004a, schreier2007a}, and even theories concerning
the origin of life\cite{roberts2018a, xu2020a} --- a detailed understanding of
these processes is therefore highly desirable.
While spectroscopic methods provide valuable insights, they face limitations in
both spatial and temporal resolution, highlighting the indispensable role of 
theoretical investigations. 
However, accurately describing such processes remains a
formidable challenge. The primary difficulty arises from the need to treat
multiple electronic states on the same footing, requiring a balanced treatment 
of both dynamical and static electron correlation, while at the same time
ensuring a feasible computational cost.

Currently, time-dependent density functional theory\cite{runge1984a} (TDDFT) 
is the most widely
used framework for excited-state calculations due to its excellent
cost-to-performance ratio. While TDDFT is, in principle, an exact 
theory,\cite{hirata1999a} 
it depends on an unknown modification of the
Hamiltonian --- specifically, the exchange-correlation potential. 
Consequently, the accuracy of TDDFT is critically determined by the choice of
exchange-correlation functional. However, given the vast array of available
approximations,\cite{mardirossian2017a} selecting an appropriate functional is 
often challenging.
Additionally, in TDDFT, excitation energies are typically computed in the
linear response regime,\cite{casida1995a} which results in an incapability to 
describe double excitations. 
The most significant limitation of TDDFT, however, is its 
inability to accurately describe static correlation effects in
the electronic ground state.\cite{matsika2021a} 
This deficiency leads to an imbalance in the
representation of the system’s electronic states, undermining its reliability
for systems where static correlation plays a major role.

The most widely used method for addressing the static correlation problem in
the ground state, thereby enabling a balanced description of different
electronic states, is the complete active space self-consistent field (CASSCF)
method\cite{roos1980a} --- the multi-reference analogue of Hartree--Fock theory. 
While CASSCF is
highly effective at capturing static correlation effects, its computational
cost increases combinatorially with the active space size and, therefore, one
is frequently restricted to small active
spaces including only the very most important orbitals. The selection of these
active orbitals is far from trivial and necessitates expert knowledge,
preventing its use as a black-box method. Often it is not even possible to
include all orbitals which should be included due to the enormous cost of the
method, resulting in a (possibly) unsatisfactory description of the system at
hand.

When considering the calculation of excited states, additional challenges
arise regarding the orbital optimization: One approach is to target the 
excited state explicitly using a
state-specific formalism. However, this method inherently biases the
calculation towards the selected state and is prone to variational collapse
onto the ground state.\cite{tran2019a} 
An alternative is the state-averaging formalism, which
optimizes a weighted average of the energies of multiple states, allowing for
simultaneous computation of several states.\cite{werner1981a}
Unfortunately, the resulting wave functions are not true stationary points on
the energy surface, which limits their accuracy compared to states obtained from
state-specific procedures.

Motivated by the aforementioned challenges, we here propose a non-orthogonal 
configuration interaction (NOCI) approach based on natural orbitals for multiple
electronic states of interest. The approach was designed with three key
objectives in mind:
(1) being systematic, (2) achieving a compact CI expansion, and 
(3) ensuring a balanced treatment of the system’s electronic states.
We will demonstrate, using simple test cases, that the presented approach forms
a promising starting point to achieve these objectives. 

\section{Non-orthogonal Configuration Interaction}
Non-orthogonal configuration interaction is a generalized extension of
the well-established orthogonal configuration interaction (CI) method. In the
NOCI framework, the generalized eigenvalue problem to determine the 
electronic states is expressed as
\begin{equation}
    \sum_{J} \left( H_{IJ} - E S_{IJ} \right) C_{J} = 0. \label{eq:NOCI}
\end{equation}
where $H_{IJ}$ and $S_{IJ}$ represent Hamiltonian and overlap matrix 
elements in the basis of Slater determinants; 
$C_{J}$ and $E$ denote the expansion coefficients and the eigenvalue 
--- the energy --- of a specific state, respectively. The
matrix elements can either be computed using L\"owdin's pairing
approach\cite{loewdin1962a, amos1961a}
in conjunction with the generalized Slater-Condon rules, or by employing the
recently introduced generalized Wick’s theorem.\cite{burton2021a}

NOCI offers significant flexibility in constructing the basis of Slater
determinants, but this flexibility also poses a key challenge: defining an
accurate and compact basis for the states of interest remains an open problem.
In this work, we propose a novel approach utilizing principal natural orbital
determinants to address this challenge.

\section{Principal Natural Orbital Determinants as NOCI Basis}

The natural orbitals (NOs) of a given electronic state are the eigenstates of 
the respective one-particle reduced density matrix (1-RDM):
\begin{equation}
    \gamma[\Psi] \ket{\phi_{p}} = n_{p} \ket{\phi_{p}}
\end{equation}
with the 1-RDM defined in a real-space basis according to
\begin{equation}
    \gamma[\Psi] = N \int \Psi^{*}(\mathbf{x}_1, \mathbf{x}_2,
    \dots) \Psi(\mathbf{x}^{'}_1, \mathbf{x}_2, \dots) 
    \mathrm{d}\mathbf{x}_2 \mathrm{d}\mathbf{x}_3 \dots
\end{equation}
The corresponding eigenvalues are the natural occupation numbers $\{n_{p}\}$ 
for which:
\begin{equation}
    n_{p} \in [0, 1]; \qquad \sum_{p} n_{p} = N_{\text{el}}
\end{equation}
where $N_{\text{el}}$ denotes the number of electrons in the system.
The NOs form an orthonormal basis and the determinant constructed from
the $N_{\text{el}}$ NOs with the highest occupation numbers is referred to as 
the principal natural orbital  determinant.\cite{yu2024a} 
In the following, we will denote the principal NO determinant 
for state $I$ (whose wavefunction is $\ket{\Psi_{I}}$) as $\ket{\Phi_{I}}$. 
Since $\gamma[\Phi_{I}]$ is the best possible 
idempotent approximation of $\gamma[\Psi_{I}]$, CI expansions based on these
determinants are highly compact.\cite{coe2015a, yu2024a}

The central idea of this work is to form, for each state of interest, a 
compact \textit{orthogonal} CI basis based on its principal NO determinant. 
Since the CI expansions of the different
states of interest are \textit{non-orthogonal} to each other, this approach
naturally leads to a NOCI problem.

To evaluate the benefits of including principal NO determinants from multiple 
states in the CI expansion --- thereby transforming the approach into a NOCI 
one --- we conducted tests on three simple systems taken from the QUEST
database.\cite{loos2018a} 
Table~\ref{tbl:effect_nat_orbs_diff_states} summarizes the energies of the two 
lowest electronic states of these systems, calculated using CISD, CASSCF, 
CASCI, and the proposed NOCI approach.
The NOCI approach, referred to as natural orbital NOCI (NO-NOCI), is annotated 
with numbers in parentheses specifying the number of principal NO determinants,
followed by the size of the active space. 
Since we aim for a compact state basis 
and double excitations should have the largest contribution 
to the correlation energy, only single and double excitations are included in
the presented NOCI approach, denoted as NO-NOCISD.
The principal NO determinants for the NO-NOCISD variants were obtained from 
preceding CASCI(4e, 12o) calculations based on PBE\cite{perdew1996a} references. 
The rational behind this decision was to lift Brillouin's Theorem to include as 
much information as possible in a CI expansion of limited length.
\begin{table}[tbp]
    \caption{Energies of the two lowest lying electronic states of the three 
    test systems formaldehyde, dinitrogen, and water obtained with various 
    electronic structure methods. 
    In all calculations, the def2-TZVP basis was employed. The NOCI methods
    employ CASCI(4e, 12o) principal NO determinants. The theorectical best
    estimate (TBE) at the complete basis set (CBS) limit was taken from
    Ref.\cite{loos2018a}}
    \label{tbl:effect_nat_orbs_diff_states}
    \begin{tabular}{lcccc}
    \hline\hline
    Method & No. Determinants & $E_{0}$ [H] & $E_{1}$ [H] & $\Delta E$ [eV] \\
    \hline
    \multicolumn{5}{c}{Formaldehyde} \\ \hline
    CISD(2e, 6o) & 36 & $-113.896215$ & $-113.753047$ & $3.90$ \\
    NO-NOCISD(1d, 2e, 6o) & 36 & $-113.926962$ & $-113.718997$ & $5.66$ \\
    NO-NOCISD(2d, 2e, 4o) & 32 & $-113.929377$ & $-113.783074$ & $3.98$ \\
    CASSCF(2e, 6o) & 36 & $-113.900280$ & $-113.811091$ & $2.43$ \\
    CASCI(4e, 12o) & 4356 & $-113.938892$ & $-113.786321$ & $4.15$ \\
    CASSCF(4e, 12o) & 4356 & $-113.993548$ & $-113.848675$ & $3.94$ \\
    \hline
    TBE(Full) CBS & -- & -- & -- & $3.58$ \\
    \hline
    \multicolumn{5}{c}{Dinitrogen} \\ \hline
    CISD(2e, 6o) & 36 & $-108.975125$ & $-108.674581$ & $8.18$ \\
    NO-NOCISD(1d, 2e, 6o) & 36 & $-109.005804$ & $-108.705817$ & $8.16$ \\
    NO-NOCISD(2d, 2e, 4o) & 32 & $-109.004870$ & $-108.727730$ & $7.54$ \\
    CASSCF(2e, 6o) & 36 & $-108.995599$ & $-108.689937$ & $8.32$ \\
    CASCI(4e, 12o) & 4356 & $-109.017840$ & $-108.730362$ & $7.82$ \\
    CASSCF(4e, 12o) & 4356 & $-109.056915$ & $-108.763409$ & $7.99$ \\
    \hline
    TBE(Full) CBS & -- & -- & -- & $7.74$ \\
    \hline
    \multicolumn{5}{c}{Water} \\ \hline
    CISD(2e, 6o) & 36 & $-76.057893$ & $-75.766529$ & $7.93$ \\
    NO-NOCISD(1d, 2e, 6o) & 36 & $-76.061898$ & $-75.464174$ & $16.26$ \\
    NO-NOCISD(2d, 2e, 4o) & 32 & $-76.061919$ & $-75.794976$ & $7.26$ \\
    CASSCF(2e, 6o) & 36 & $-76.052454$ & $-75.815680$ & $6.44$ \\
    CASCI(4e, 12o) & 4356 & $-76.085993$ & $-75.803505$ & $7.69$ \\
    CASSCF(4e, 12o) & 4356 & $-76.108827$ & $-75.858557$ & $6.81$ \\
    \hline
    TBE(Full) CBS & -- & -- & -- & $7.33$ \\
    \hline\hline
    \end{tabular}
\end{table}

Inspecting Table~\ref{tbl:effect_nat_orbs_diff_states}, several observations 
are particularly noteworthy. Most importantly, when 
comparing the NOCI approach using only the principal NO determinant of the 
ground state with the approach including the principal NO determinants of both 
the ground and first excited states, the latter provides a significantly 
improved description of the first excited state. 
Table~\ref{tbl:performance_water} further illustrates the impact of the initial 
guess quality and the size of the single and double excitation space on the
performance of the NO-NOCISD(1d) approach. As the quality of the initial guess
increases, the NOs become more and more tailored to the state they were obtained 
for --- in this case, the ground state. 
However, this results in a progressively poorer description of the first 
excited state. Expanding the excitation space in the NO-NOCISD(1d) approach 
significantly enhances the description of the excited state, while the 
improvement for the ground state is comparatively moderate.
This nicely demonstrates the ordering of the NOs according to their relevance 
for describing the electronic state they were obtained for. Moreover, it
underscores the necessity of including the principal NO determinants for
\textit{all} states of interest within the NOCI framework to achieve a compact
and accurate basis.
\begin{table}[tbp]
    \caption{Impact of the initial guess quality and the size of the 
    excitation space on the performance of the NO-NOCISD(1d) approach. 
    All calculations were performed with the def2-TZVP basis set.}
    \label{tbl:performance_water}
    \begin{tabular}{lcc}
    \hline\hline
    Method & $E_{0}$ [H] & $E_{1}$ [H] \\
    \hline
    \multicolumn{3}{c}{CASCI(2e, 4o) Initial Guess} \\ \hline
    NO-NOCISD(1d, 2e, 6o)  & $-76.057894$ & $-75.766529$ \\
    NO-NOCISD(1d, 2e, 8o)  & $-76.058190$ & $-75.768593$ \\
    NO-NOCISD(1d, 2e, 10o) & $-76.058768$ & $-75.771224$ \\
    \hline
    \multicolumn{3}{c}{CASCI(4e, 12o) Initial Guess} \\ \hline
    NO-NOCISD(1d, 2e, 6o)  & $-76.061898$ & $-75.464174$ \\
    NO-NOCISD(1d, 2e, 8o)  & $-76.062828$ & $-75.516464$ \\
    NO-NOCISD(1d, 2e, 10o) & $-76.062859$ & $-75.685024$ \\
    \hline
    \multicolumn{3}{c}{CASCI(6e, 14o) Initial Guess} \\ \hline
    NO-NOCISD(1d, 2e, 6o)  & $-76.066963$ & $-75.360981$ \\
    NO-NOCISD(1d, 2e, 8o)  & $-76.067515$ & $-75.365945$ \\
    NO-NOCISD(1d, 2e, 10o) & $-76.067968$ & $-75.401444$ \\
    \hline\hline
    \end{tabular}
\end{table}

When comparing NO-NOCISD(2d, 2e, 4o) with CISD using a similar number of 
determinants, NO-NOCISD(2d, 2e, 4o) shows significantly higher accuracy for the 
absolute energies of the two states, although CISD performs well in describing 
their energy difference. 
Similarly, when comparing NO-NOCISD(2d, 2e, 4o) with CASSCF employing a 
comparable number of determinants, the NOCI approach demonstrates 
overall better performance, particularly in reproducing the energy difference 
between the states.

It is also noteworthy that NO-NOCISD(2d, 2e, 4o) recovers the energies obtained
with CASCI(4e, 12o) --- which was used to generate the principal NO 
determinants --- to a very good extent while utilising only about 1\% of the 
determinants required by CASCI. To be more explicit, the worst agreement is
observed for $E_0$ of water, where NO-NOCISD(2d, 2e, 4o) still recovers 99.97\%
of the total CASSCI(4e, 12o) energy.

In summary, incorporating the principal NO determinants of multiple states 
proves highly advantageous. It enables the construction of a compact NOCI basis 
that achieves comparable accuracy across the states of interest.

\section{Orthogonal Determinant Selection}

Thanks to the orthogonality ``within'' each state of interest, we can compute 
the corresponding Hamiltonian elements more efficiently and further leverage 
established algorithms to identify the most significant determinants.

Adopting the strategy used in heat-bath CI\cite{holmes2016a,holmes2016b}, we 
include, for each state of interest, only the singly and doubly excited 
determinants that are connected to the respective principal NO determinant by a 
Hamiltonian matrix element with a magnitude exceeding a specified threshold 
$\epsilon$.
Since these determinants are othogonal to each other, the respective matrix
elements of the Hamiltonian can be evaluated according to
\begin{equation}
    H(r \leftarrow p) = \Gamma_{rp}^I \left( h_{rp} 
    + \sum_{q \in \text{occ}} \left( g_{rqpq} - g_{qrpq} \right) \right)
\end{equation}

\begin{equation}
    H(rs \leftarrow pq) = \Gamma_{rp}^I \Gamma_{sq}^J 
    \left( g_{rspq} - g_{srpq} \right)
\end{equation}
with
\begin{align}
    h_{rp} &= \int \phi_r^*(\mathbf{x}) \left( -\frac{1}{2} \nabla^2 - 
    \sum_A \frac{Z_A}{|\mathbf{r} - \mathbf{R}_A|} \right) 
    \phi_p(\mathbf{x}) \, \textrm{d}\mathbf{x} \\
    g_{rspq} &= \int \phi_r^*(\mathbf{x}_1) \phi_s^*(\mathbf{x}_2) 
    \frac{1}{|\mathbf{r}_1 - \mathbf{r}_2|} \phi_p(\mathbf{x}_1) 
    \phi_q(\mathbf{x}_2) \, \textrm{d}\mathbf{x}_1 \, \textrm{d}\mathbf{x}_2
\end{align}
and $\Gamma_{rp}^I = (-1)^n$, where $n$ is the number of occupied spin-orbitals
between $p$ and $r$ in state $I$.
As we are concerned only with the magnitude of the matrix elements and not 
their sign, double excitations can be handled very efficiently. This is because 
the magnitude depends solely on the four orbitals whose occupations change 
during the excitation and not the other orbitals.

\begin{figure}[htbp]
    \centering
    \begin{subfigure}{0.48\textwidth}
        \centering
        \includegraphics[width=\textwidth]{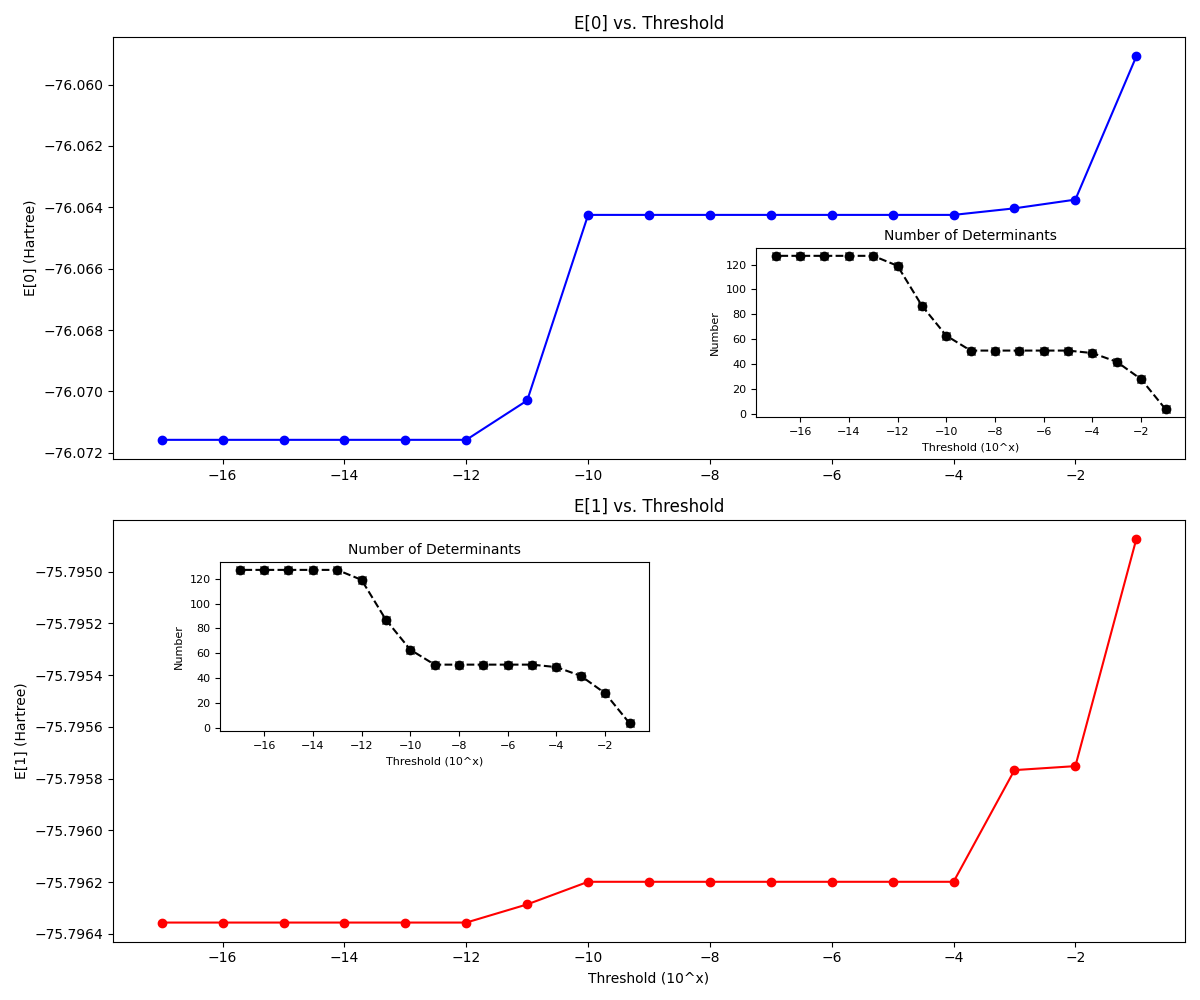}
        \caption{NO-NOCISD(2d, 2e, 8o).}
        \label{fig:water_det_sel_2d2e8o}
    \end{subfigure}
    \hfill
    \begin{subfigure}{0.48\textwidth}
        \centering
        \includegraphics[width=\textwidth]{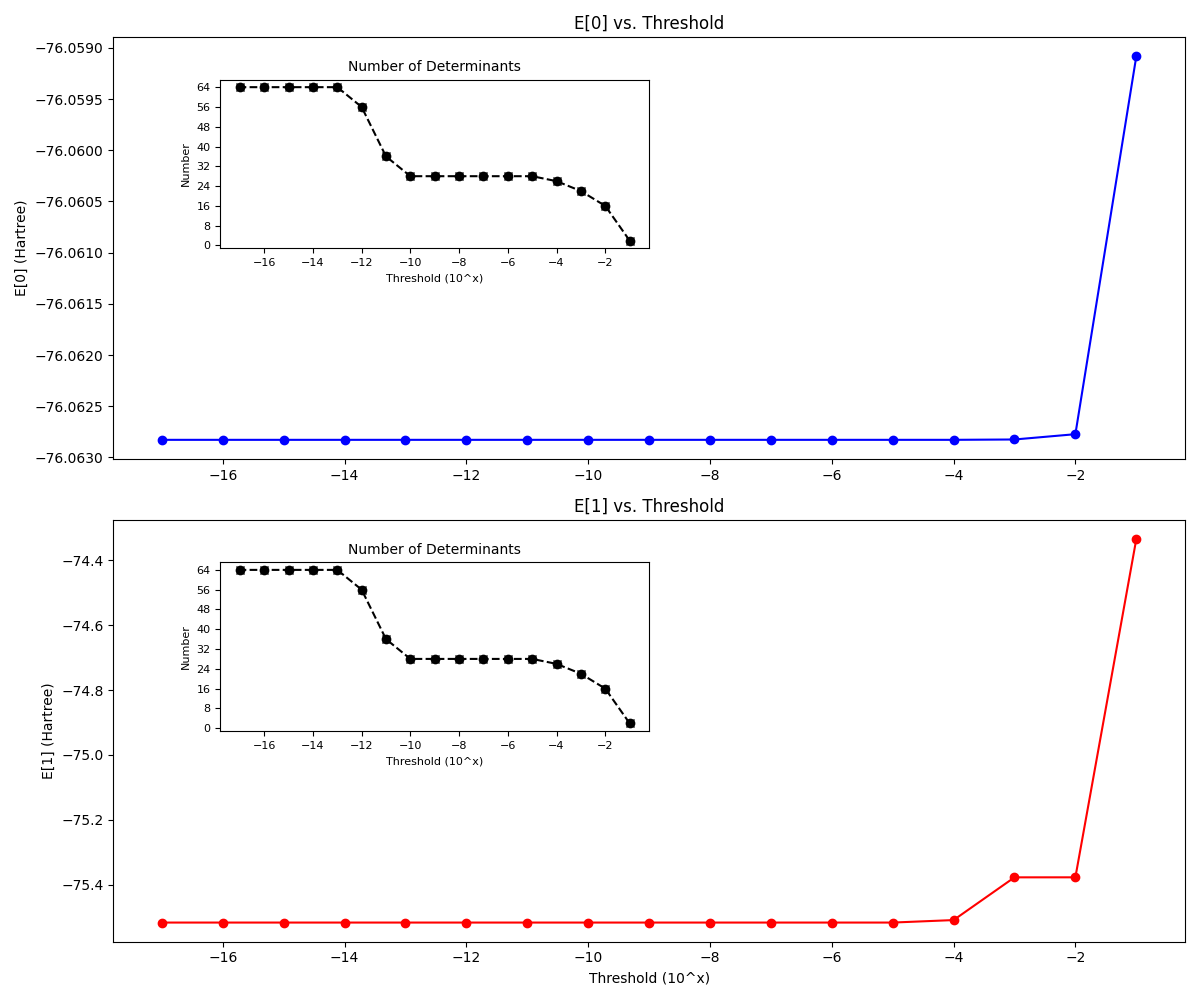}
        \caption{NO-NOCISD(1d, 2e, 8o).}
        \label{fig:water_det_sel_1d2e8o}
    \end{subfigure}
    \caption{Energies of the ground and the first excited state of \ce{H2O}
    evaluated with two different NO-NOCISD variants in the zeroth iteration.
    The def2-TZVP basis set was employed. 
    The initial guess was obtained with CASCI(4e, 12o).}
    \label{fig:water_det_comparison}
\end{figure}

To assess the impact of the determinant selection, we conducted tests on the 
water molecule using the def2-TZVP basis set. A comparison of
Fig.\ref{fig:water_det_sel_2d2e8o} and Fig.\ref{fig:water_det_sel_1d2e8o}
reveals a significant improvement in the energies of both the ground and first
excited states for small selection thresholds in the case of NO-NOCISD(2d, 2e,
8o). This improvement can be attributed to the non-orthogonal mixing of
determinants from the two states, as indicated by the simultaneous
energy jumps observed in both states, which are absent in the NO-NOCISD(1d, 2e,
8o) case.

These results suggest a critical role of small Hamiltonian matrix elements
in enhancing the ability of the NOCI approach to improve state descriptions 
through non-orthogonal mixing. 
Incorporating non-orthogonal information into the selection process could 
further improve the accuracy; however, this would entail
significantly higher computational costs. For the sake of efficiency, we will
adhere to the orthogonal approach outlined above in this work.

\section{Iteratively Recalculating the Principal NO Determinant}
To further reduce the size of the CI expansion, we tested an iterative process
that involves repeatedly computing the 1-RDMs of the states of interest,
forming the corresponding principal NO determinants, and selecting the most
important determinants within a predefined active space using the procedure
described earlier. 
The underlying idea is to leverage the non-orthogonal flexibility of the
framework to encode information during the orbital optimization process,
ultimately resulting in a more compact state basis.
A further advantage of this approach is its potential to reduce dependence on 
the initial guess, which has thus far played a critical role since the 
principal NO determinants were obtained from the initial guesses for the 1-RDMs 
of the states of interest. 
While the influence of the initial guesses naturally decreases with larger
active excitation spaces within the NOCI framework, the presented approach 
prioritizes maintaining a compact CI expansion, making small active spaces 
preferable.

For our investigations we again used water as test system and focused on
two questions:
1) Does the iterative procedure make the CI expansion more compact? 
2) Does it affect the accuracy?

\begin{figure}[htbp]
    \centering
    \begin{subfigure}{0.48\textwidth}
        \centering
        \includegraphics[width=\textwidth]{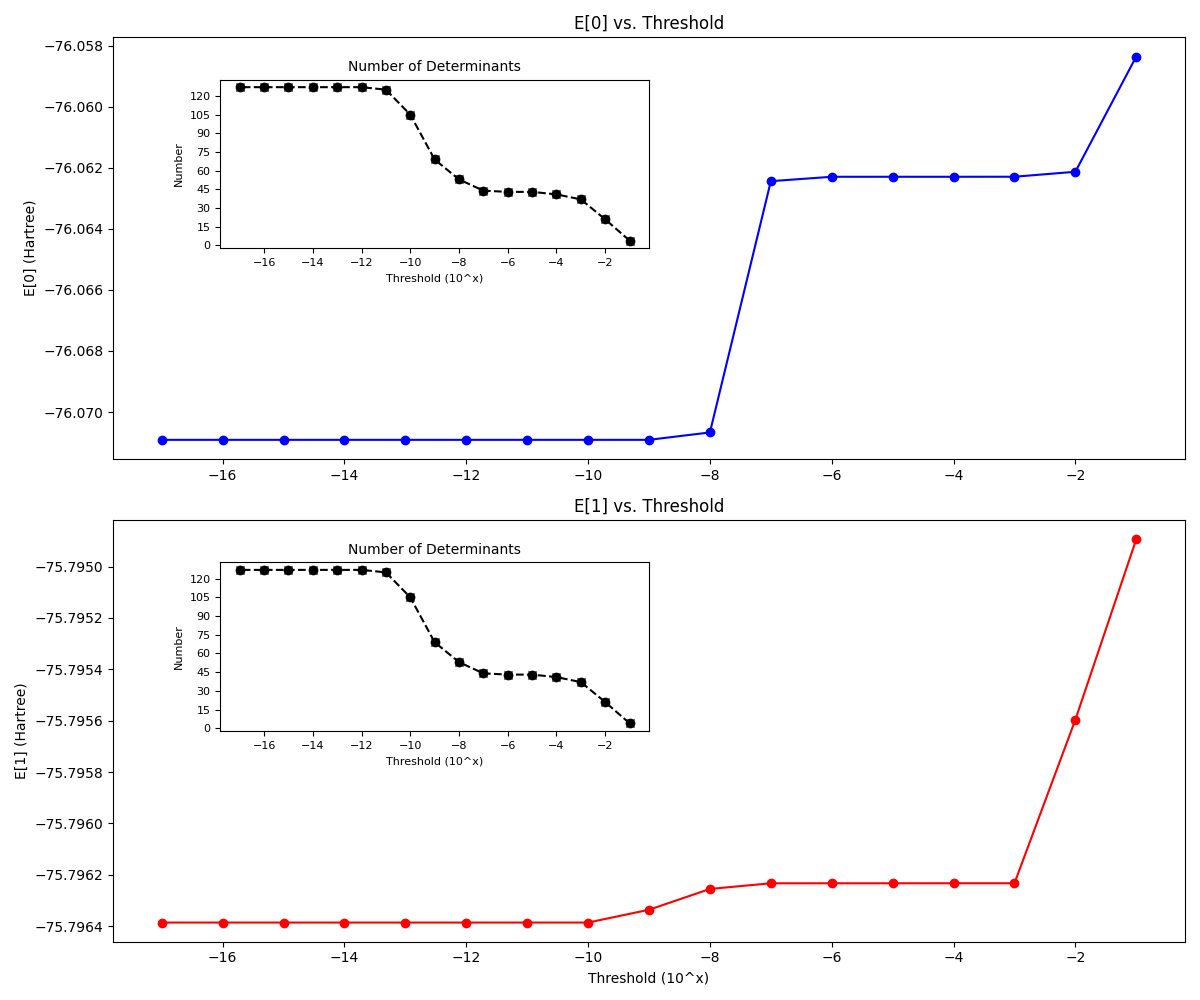}
        \caption{NO-NOCISD(2d, 2e, 8o).}
        \label{fig:water_det_sel_2d2e8o_it1}
    \end{subfigure}
    \hfill
    \begin{subfigure}{0.48\textwidth}
        \centering
        \includegraphics[width=\textwidth]{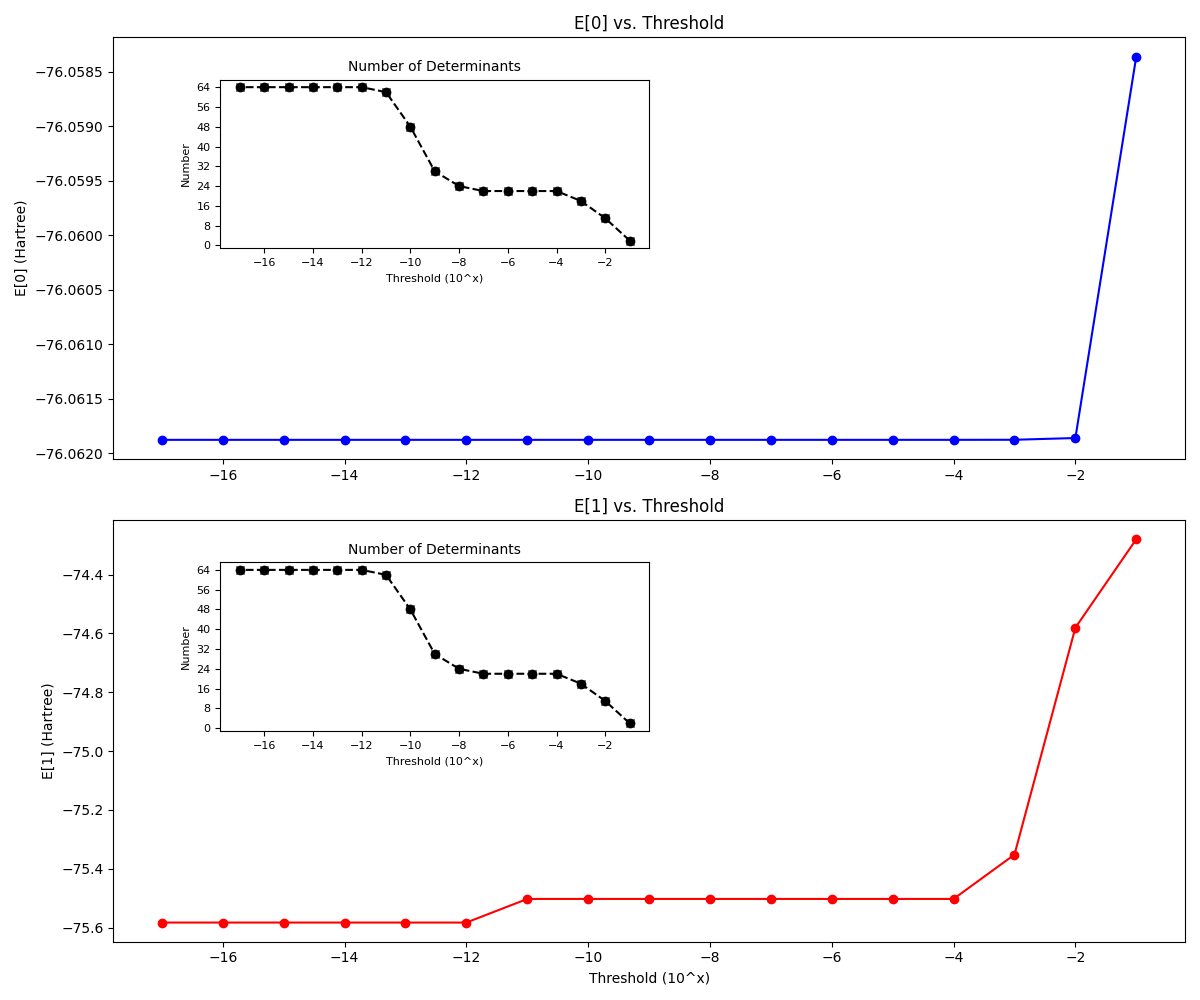}
        \caption{NO-NOCISD(1d, 2e, 8o).}
        \label{fig:water_det_sel_1d2e8o_it1}
    \end{subfigure}
    \caption{Energies of the ground and the first excited state of \ce{H2O}
    evaluated with two different NO-NOCISD variants in the first iteration.
    The def2-TZVP basis set was employed. 
    The initial guess was obtained with CASCI(4e, 12o).}
    \label{fig:water_det_comparison_it1}
\end{figure}

\begin{figure}[htbp]
    \centering
    \begin{subfigure}{0.48\textwidth}
        \centering
        \includegraphics[width=\textwidth]{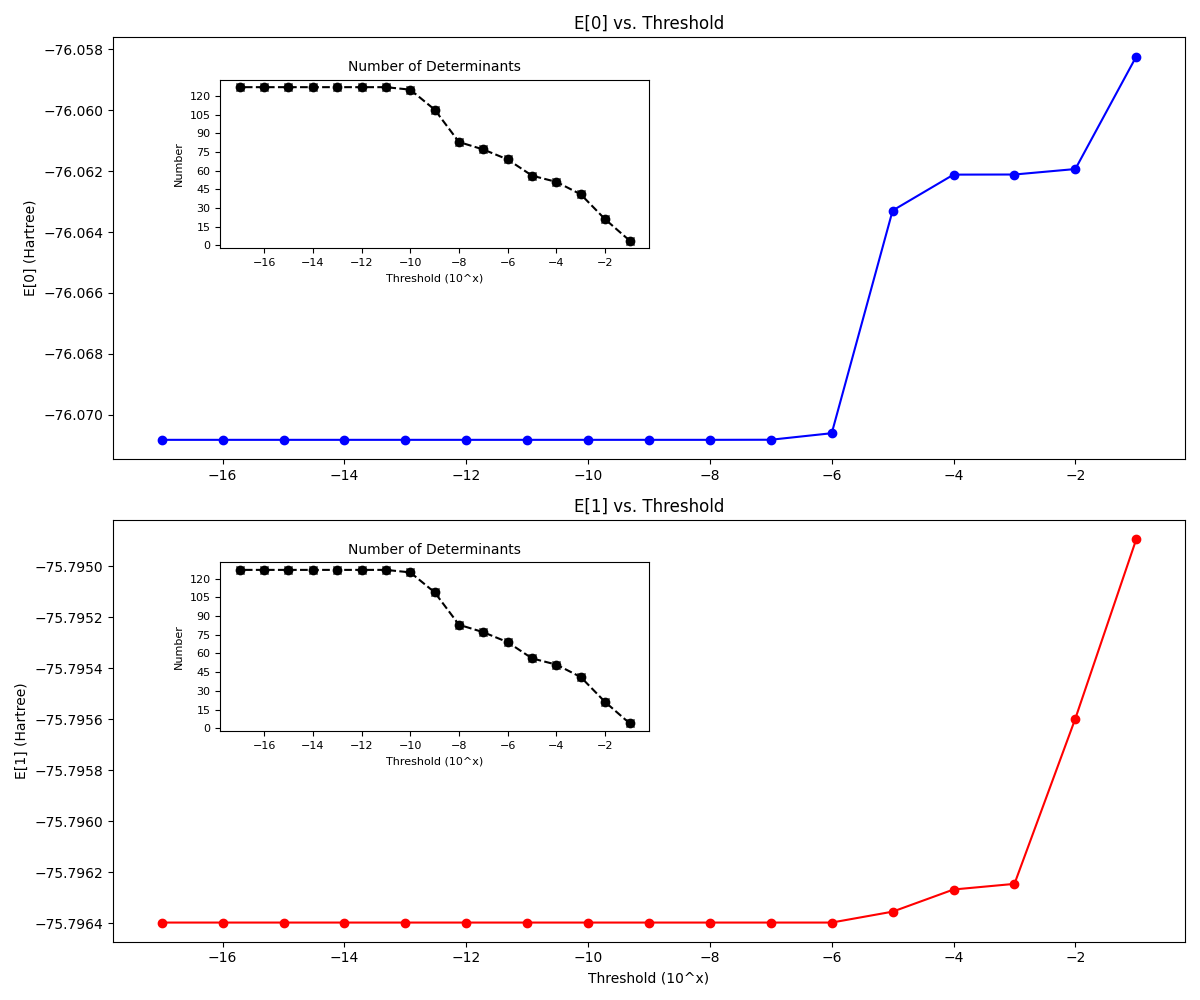}
        \caption{NO-NOCISD(2d, 2e, 8o).}
        \label{fig:water_det_sel_2d2e8o_it2}
    \end{subfigure}
    \hfill
    \begin{subfigure}{0.48\textwidth}
        \centering
        \includegraphics[width=\textwidth]{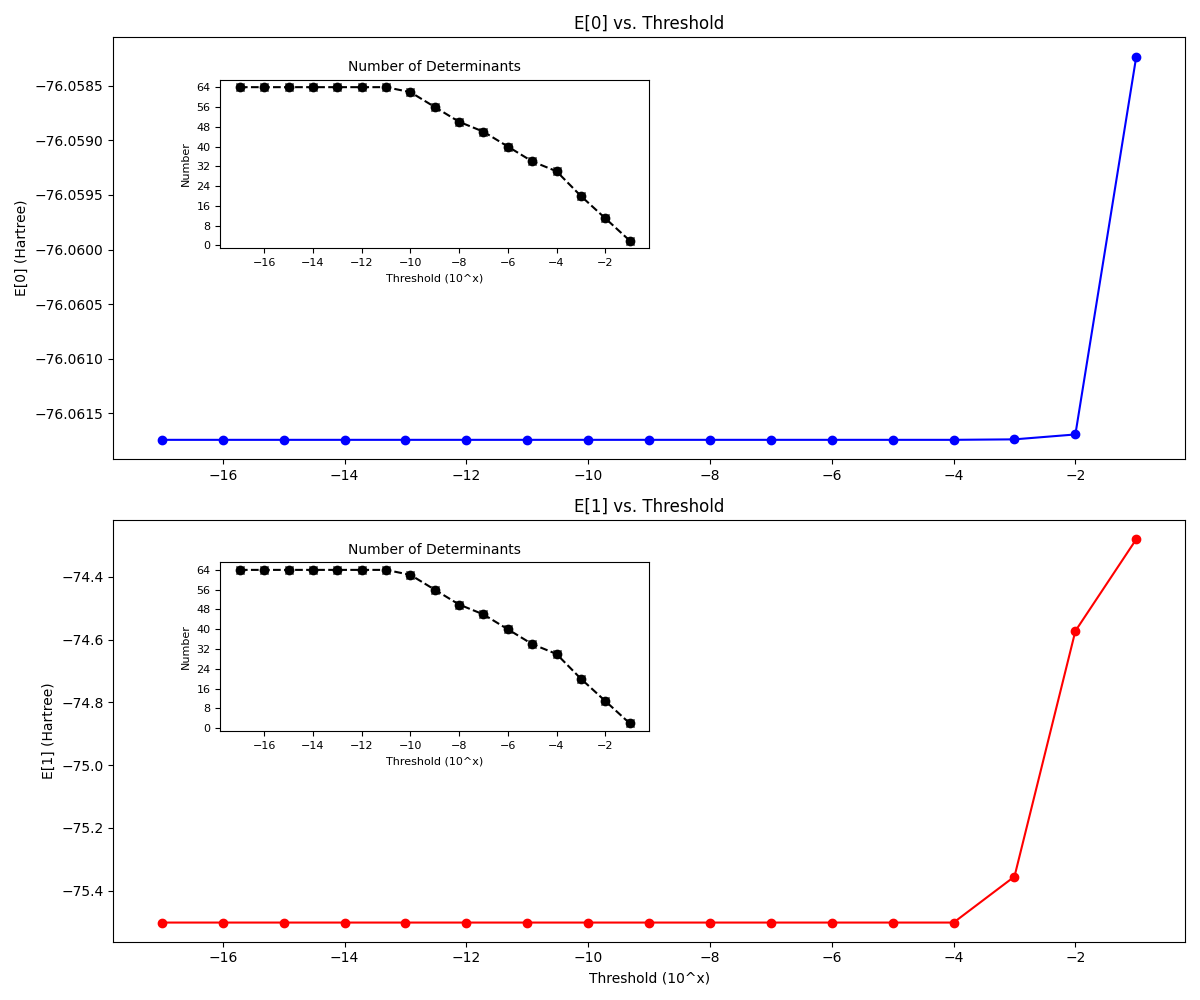}
        \caption{NO-NOCISD(1d, 2e, 8o).}
        \label{fig:water_det_sel_1d2e8o_it2}
    \end{subfigure}
    \caption{Energies of the ground and the first excited state of \ce{H2O}
    evaluated with two different NO-NOCISD variants in the second iteration.
    The def2-TZVP basis set was employed. 
    The initial guess was obtained with CASCI(4e, 12o).}
    \label{fig:water_det_comparison_it2}
\end{figure}

1) Fig.~\ref{fig:water_det_comparison_it1} and
Fig.~\ref{fig:water_det_comparison_it2} again show the energies of the ground
and the first excited states as functions of the determinant selection
threshold; this time, however, for iteration 1 and 2, respectively. 
Notably, the threshold at which the determinants of the two states begin to mix
increases with each iteration. Additionally, fewer determinants are needed to
achieve comparable accuracy (see inset). These results hence suggest that the
NOCI expansion size can indeed be reduced by the employed iterative
procedure.

2) Tab.~\ref{tab:energies_iterative_proc} summarizes the energies of the ground
and first excited states as a function of the iteration, comparing a less
accurate initial guess (left) with a significantly better initial guess
(right). Two key observations can be made: First, for the less accurate initial
guess, the iterative procedure leads to notable improvements in accuracy.
Second, for the better initial guess, the iterative procedure does not enhance 
the accuracy but also does not lead to any significant deterioration.
These findings suggest that the iterative procedure is particularly beneficial
when starting from poor initial guesses, while posing low risk to accuracy
when the initial guess is already good. Naturally, larger active excitation
spaces within the NOCI framework increase the potential for energy improvements
during the iterative process and thereby further reduce the dependence on the
initial guesses; however, since this approach prioritizes
compactness, we intentionally limit the size of the active spaces.
\begin{table}[htbp]
    \centering
    \begin{tabular}{@{}cccccc@{}}
    \toprule
    Iteration & \multicolumn{2}{c}{CASCI(2e, 4o) Initial Guess} &
    \multicolumn{2}{c}{CASCI(4e, 12o) Initial Guess} \\
    \cmidrule(lr){2-3} \cmidrule(lr){4-5}
              & $E_0$ & $E_1$ & $E_0$ & $E_1$ \\
    \midrule
    0 & $-76.058190$ & $-75.768593$ & $-76.071585$ & $-75.796357$ \\
    1 & $-76.060309$ & $-75.769467$ & $-76.070916$ & $-75.796386$ \\
    2 & $-76.061190$ & $-75.769925$ & $-76.070839$ & $-75.796397$ \\
    3 & $-76.061207$ & $-75.769931$ & $-76.070801$ & $-75.796402$ \\
    4 & $-76.061205$ & $-75.769932$ & $-76.070778$ & $-75.796410$ \\
    \bottomrule
    \end{tabular}
    \caption{NO-NOCISD(2d, 2e, 8o) energies of the ground and first excited 
    states of \ce{H2O} across
    different iterations, starting from two initial guesses of different
    quality. The def2-TZVP basis set was employed.}
    \label{tab:energies_iterative_proc}
\end{table}

Regarding initial guesses for the proposed approach, it would be valuable to 
further explore methods other than CASSCI. Particularly
interesting are methods that 1) treat different states on an equal footing and
2) are computationally inexpensive. Potential candidates include spin-flip
TDDFT\cite{shao2003a, orms2018a, krylov2017a, bernard2012a}, the
constraint-based orbital-optimized excitation (COOX)
method\cite{kussmann2024a}, and the maximum overlap method.\cite{gilbert2008a}
However, investigations in these directions are left for future work.

\section{Final Procedure}
Based on the findings of the preceding sections, we propose the following
procedure for constructing a compact NOCI basis:

\begin{enumerate}[itemsep=0pt, parsep=0pt]
    \item Define the states of interest (number of roots).
    \item Compute initial guesses for the 1-RDMs of the states of interest.
    \item Construct the principal NO determinant for each state of interest 
        and include them in the NOCI expansion.
    \item For each principal NO determinant, generate and include only those 
        singly and doubly excited determinants within a predefined active space 
        that are connected to the respective principal NO determinant by a 
        Hamiltonian matrix element exceeding an adaptive threshold $\epsilon$
        in magnitude.
    \begin{itemize}
        \item Adjust $\epsilon$ to account for the increasing magnitude of the 
            Hamiltonian matrix elements during the iterative process and to
            ensure that the NOCI expansion size is reduced.
    \end{itemize}
    \item Solve the generalized eigenvalue problem defined in Eq.~\ref{eq:NOCI}.
    \item Recalculate the 1-RDMs based on the updated NOCI states.
    \item Repeat steps 3 through 6 until the degradation in the energy values
        of the states of interest exceeds a predefined threshold.
\end{enumerate}

\section{Computational Details}
The NO-NOCI approach was implemented in an in-house Python program interfaced
with the FermiONs++ program package.\cite{kussmann2013a, kussmann2015a} All
CISD and CASCI calculations were carried out using FermiONs++, while CASSCF
calculations were performed with the ORCA quantum chemistry program
package.\cite{neese2020a, neese2022a, kollmar2019a, neese2023a, ugandi2023a}
The def2-TZVP basis set\cite{weigend2005a} was used consistently for all
calculations.

\section{Conclusion}

In this work, we introduced a novel procedure for constructing a compact NOCI
basis to efficiently treat ground and excited states. The
core idea is to include the principal natural orbital determinant of each state
of interest in the NOCI expansion, ensuring all states are treated on equal
footing. To maintain compactness, only the most important single and double
excitations from these reference NO determinants are added to the expansion.
The inclusion of multiple NO determinants naturally leads to a NOCI framework;
however, the orthogonality ``within'' each state allows for efficient 
calculation of Hamiltonian matrix elements and determinant selection.

We employ an iterative procedure where the 1-RDMs of all states of interest are
recalculated at each step to update the principal NO determinants. This
approach fully exploits the flexibility of the non-orthogonal framework,
enabling a further reduction in the expansion size without significant
deterioration of accuracy.

Overall, this work represents a promising step toward a systematic procedure for
generating compact and accurate NOCI expansions. However, we emphasize that
this study serves as a proof of concept, and further testing and refinement
will be needed. We hope that this work inspires further exploration and
development in this area.

\begin{acknowledgement}
    D.~G.~acknowledges funding by the Deutsche Forschungsgemeinschaft 
    (DFG, German Research Foundation) -- 498448112. 
    D.~G.~thanks J.~Kussmann (LMU Munich) for providing a development version 
    of the FermiONs++ software package.
\end{acknowledgement}

\bibliography{lit}

\end{document}